\documentstyle[aps, pra, multicol, epsfig]{revtex}
\def\bea{\begin{eqnarray}}
\def\eea{\end{eqnarray}}
\def\be{\begin{equation}}
\def\ee{\end{equation}}
\begin{document}
\draft
\author{Bogdan Damski,  Krzysztof Sacha, 
and Jakub Zakrzewski}
\address{
Instytut Fizyki imienia Mariana Smoluchowskiego,
  Uniwersytet Jagiello\'nski,\\
 ulica Reymonta 4, PL-30-059 Krak\'ow, Poland \\
}
\title{Collective  excitation of trapped degenerate Fermi gases}
\date{\today}
\maketitle
\begin{abstract}
We show that the slow driving of a focused laser beam through the cloud of 
trapped cold fermions allows for a creation of the collective excitation 
in the system. 
The method, proposed originally by us for bosons, seems to be quite 
feasible experimentally --- it requires only a proper change in time of 
the potential in atomic traps, as realized 
in laboratories already.
\end{abstract}
\begin{multicols}{2}

The experimental realization of the Bose-Einstein condensate (BEC) \cite{cornel} 
in a trapped Bose gas has triggered considerable interest in the field of 
cold degenerate atomic gases \cite{parkins98,dalfovo99}.
The relatively weak interaction between atoms allows for precise experimental 
manipulation of an atomic gas. In particular collective excitations of the BEC
such as vortices and solitons \cite{cornell99,dalibard,phillips,Hannover}
(in the mean field language -- see also \cite{jacek}) 
have been studied in details both 
theoretically and experimentally. 
Nowadays, it is also possible to cool down Fermi gases to a regime 
where effects of the quantum statistics become noticeable 
\cite{demarco99,schreck01,truscott01}.
Therefore, it would be interesting to contrast
their collective behaviors \cite{vichi,clark,karpiuk01} with those of the BEC. 

Very recently we have proposed a new simple scheme that allows for creation 
of both solitons and vortices in a BEC using an appropriate time
dependent modification (by an additional tightly focused laser beam)
of the trapping potential \cite{kark01,bodzio01}. The aim of the present 
work is to discuss possible collective excitations of trapped fermions and  
to show that the very same scheme as for bosons may be utilized successfully to
create excitation in a cold Fermi gas. A numerical attempt 
to create such excitations using the phase imprinting method 
\cite{dobrek99}, has been recently made \cite{karpiuk01}.
It is not clear, however, which states are really excited in this scheme.

The dominant, at low temperatures, $s$-wave collisions 
are prohibited for spin-polarized
identical fermions due to the Pauli exclusion principle.
Consequently, it is natural to consider first a noninteracting particle model 
as the atoms may interact only through vanishingly small $p$-wave collisions. 
Our excitation method originally proposed for noninteracting bosons remains valid even for 
quite strong interaction between particles. We expect thus that consideration
of the noninteracting Fermi system as a first approximation is perfectly
legitimate. 

It is well known that BCS transition may occur in the quantum cold
gas of fermions at a very low temperature $T_c$ \cite{stoof,you99}. 
While in the noninteracting particle model this effect is absent,
we shall show that creation of collective excitations may occur both for 
$T=0$ (neglecting BCS pairing) and for temperatures that are greater 
than $T_c$ but sufficiently low that the effects of quantum statistics 
are noticeable. 

The paper is organized as follows. Firstly we specify what we consider
as a collective
excitation in the cold fermionic gas on a one-dimensional (1D)
example. Secondly, we present the basic idea of the excitation scheme
still for 1D
noninteracting particle model at zero temperature limit.
Finally we move to 
the 3D model for an experimentally realistic temperature 
indicating that signatures of the collective excitations can be observed
in a laboratory. 

A gas of bosons being in a product state of ground, single
particle (e.g. mean field) states is a standard approximate description
of the BEC. It is natural then to consider as a collective excitation a
situation in which all particles are simultaneously transferred to
some excited state of the mean field effective potential. This results in 
the so called ``dark soliton'' \cite{dum98}. Since all particles are 
both initially and finally in the same state, the mean field description
of the soliton creation requires an efficient mechanism for transfer
the population from the ground to the first excited state.

\begin{figure}
\centering
{\epsfig{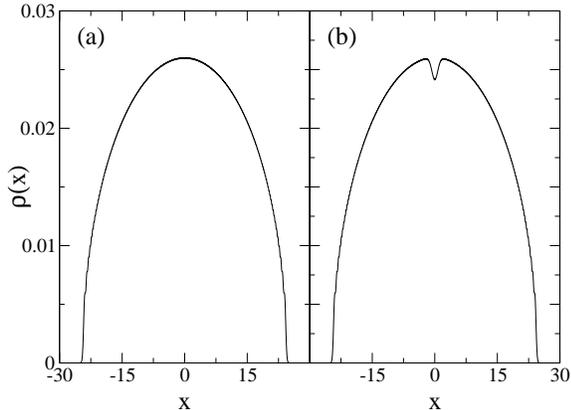}}
\caption{
Single particle reduced probability density for $N = 300$ fermions in 
the 1D model,
at temperature $T=0$, before [panel (a) -ground state] and after 
[panel (b)- collectively excited state] the sweeping of 
the perturbation. For $x$ we use the harmonic oscillator unit of length, i.e. 
$\sqrt{\hbar/m\omega}$.
}
\label{ideal}
\end{figure}

The situation is quite different for Fermi system for which each state can be
at most single occupied (we assume a spin-polarized identical fermions as realized 
experimentally).
Consider a 1D  Fermi gas at $T=0$ prepared in 
an ideal collective ground state of the harmonic trap. 
For $N$ identical fermions this is
equivalent to assuming that all oscillator eigenstates, $\psi_n(x)$, from $n=0$
(single particle ground state)  to
$n=N-1$ are occupied with other levels being necessarily empty.
The wave function of the $N$ fermion system 
may be represented then by the Slater determinant
\begin{eqnarray} 
\label{slater}
\Psi(x_1, x_2, ..., x_N)=& \cr
\frac{1}{\sqrt{N!}}& \left|
\begin{array}{cccc}
\psi_0(x_1)&\psi_0(x_2)  & \cdots  & \psi_0(x_N)  \\
\vdots & \vdots &  & \vdots   \\
\psi_{N-1}(x_1) & \psi_{N-1}(x_2) & \cdots & \psi_{N-1}(x_N)
\end{array} \right|.
\end{eqnarray}
The corresponding single particle reduced probability density reads
\[
\varrho(x) = \int dx_2...dx_N |\Psi(x, x_2, ... , x_N)|^2 
= \frac{1}{N} \sum_{i=0}^{N-1} |\psi_i(x)|^2
\]
which (employing Christoffel-Darboux formula  \cite{atlas}) may be reduced to
\begin{eqnarray}
\varrho(x) & = & \frac{\exp(-x^2)}{\sqrt{\pi} N! 2^{N-1}} 
[N H_{N-1}^2(x)  \nonumber \\ && -  
(N-1)H_N(x)H_{N-2}(x)],
\label{density}
\end{eqnarray}
where $H_N(x)$ denote standard Hermite polynomials.
This density is shown in Fig.~\ref{ideal}a.

Now since each fermion occupies a different state, there is some ambiguity in 
defining a collective excitation. We define it by requiring that
each fermion undergoes a single excitation. Since they are indistinguishable,
the final wave function is
then the Slater determinant involving the oscillator functions from
$n=1$ to $n=N$, the corresponding single particle density 
is shown in Fig.~\ref{ideal}b. This is a most
complete analogy to the collective excitation of bosons where all $N$
particles gain a single excitation too. The difference is that bosons
enjoy the same initial (and final) state while the fermions need to be
excited from necessarily different states. Importantly, all bosons
pass from an even to an odd state in an ideal case, leading to the creation
of a ``dark soliton'', with a node in the center. 
The collective excitation of a fermionic gas leaves the ground state empty 
that results in a dip in the single particle probability density 
(Fig.~\ref{ideal}b). In the 1D model the relative depth of the dip 
scales with number of fermions as $1/\sqrt{N}$.

How to realize such a final ``collective'' state? 
Surprisingly we show below that a
patient way of exciting fermions one by one, is quite simple and feasible.
Moreover, it turns out that in a practical implementation it seems to be
an almost trivial extension of the scheme we proposed for an efficient
excitation of a Bose gas!

As in the original suggestion for the bosonic case \cite{kark01} we propose
to modify the harmonic trapping potential by an additional tightly focused 
laser beam which we sweep through the trap. Provided that the frequency 
of this laser beam is  sufficiently detuned from an internal atomic
transition (still being close to the resonance so that two
atomic levels can be considered only) the upper level may be eliminated
adiabatically from the analysis (see e.g. \cite{zoller}). Then the 
effective potential for the atomic motion, in the 1D
model, reads
\begin{equation}
V(x)=\frac{x^2}{2} + U_0(x_0) 
\exp\left(\frac{-(x-x_0)^2}{2\sigma^2}\right),
\label{potencja}
\end{equation}
where $U_0(x_0)<0$ (for an appropriately chosen laser detuning).
 The amplitude $U_0(x_0)$ is proportional to the 
laser intensity, $x_0=x_0(t)$ is a time-dependent position of the center of 
the laser beam, while $\sigma$ is directly related to the cross-section of 
the (gaussian) laser beam.
We assume the trapping harmonic oscillator units, i.e.
$\omega t$ for time and $\sqrt{\hbar/m\omega}$ for length, 
 where $m$ stands for an atomic mass.

Assume we prepare the fermionic sample in the state (\ref{slater}) in a pure
harmonic trap.
Next we turn on the laser, initially focused at the left edge 
of the trap  ($x_0(0)\ll 0$) and move the focus across
the trap towards the center realizing the potential (\ref{potencja}).
For all numerical simulations presented below we take
$\sigma = 0.2$ and 
\begin{equation}
U_0(x_0) = (a-b x_0) \arctan(x_0),
\label{potla}
\end{equation}
with $a,b>0$. 
In the bosonic case  we took $U_0(x_0)$ proportional to 
$\arctan(x_0)$ \cite{kark01,bodzio01}. 
In the present case we have to add an additional prefactor
ensuring that the perturbation is not negligible in comparison with the  
harmonic part (which is large far from the center of the trap).
 We have chosen the simplest linear form of such a prefactor
taking $a=18$, $b=0.5$ in the numerical examples,
the results are not very sensitive to precise values of $a$ and $b$. 
During the excitation process
$x_0$ changes linearly with time according to $x_0(t)=x_0(0)+0.02  t$.

\begin{figure}
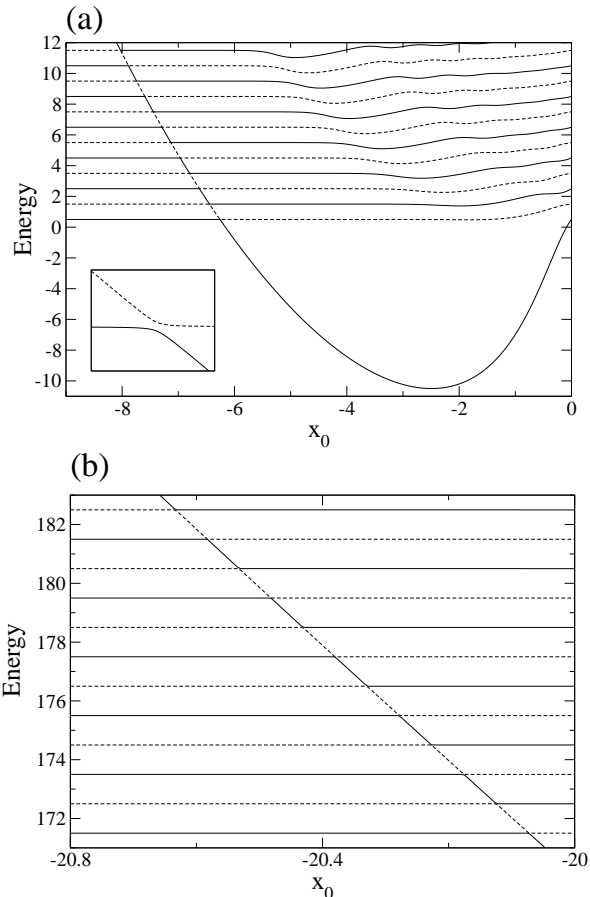

\centering
{\epsfig{file=damski02a.eps, clip = true, width=7.5cm}}
{\epsfig{file=damski02b.eps, clip = true, width=7.8cm}}
\caption{
Energy levels of the potential (\ref{potencja}) as a function of $x_0$. 
Panel (a) shows the behavior of low lying energy levels while 
in the panel (b) the energy levels around $n=180$ are presented.
Note a series of very narrow avoided crossings between the neighboring energy
levels (for clarity consecutive levels are drawn using solid and dashed lines).
The inset in panel (a) shows a vastly enlarged avoided crossing between the ground
and the first excited state around $x_0\approx -6.2$. In the figure we use 
the harmonic oscillator units, i.e. $\hbar\omega$ for energy and 
$\sqrt{\hbar/m\omega}$ for length.
}
\label{poz1}
\end{figure}

 To understand the effect produced by sweeping of the laser induced
 potential across the trap
it is sufficient to consider the change of
single particle energy levels as a function of the center of
the laser beam $x_0$. The corresponding level dynamics is depicted
in Fig.~\ref{poz1}. Observe that the levels undergo a series
of orderly arranged very narrow avoided crossings 
(not only between low lying levels but also between the highest ones).  
Assume 
$x_0$ is changed sufficiently slowly so as to follow the energy levels
adiabatically except in the immediate vicinities of avoided crossings.
Here, since the avoided crossings are extremely narrow we assume that
they are passed diabatically. Under these premises the excitation 
scenario works as follows. We start with the laser beam situated at
negative value of $x_0$ (with large $|x_0|$) and increase $x_0$ up to 
$x_0=0$ where the sweeping process ends (when the center of the laser beam
reaches the center of the harmonic trap its intensity is reduced to zero).
With increasing $x_0$ we pass diabatically 
the consecutive avoided crossings encountered on the way. The important
avoided crossings occur when one of the levels is occupied. Assuming
that the highest occupied level is $N-1$, the diabatic passage via
the avoided crossing of the $N$ with the $N-1$ level populates the $N$
level leaving practically empty the $N-1$ state (the Landau-Zener effect). 
The next crossing,
occurring at slightly larger $x_0$ value populates now  the empty $N-1$
level from $N-2$,  another one populates $N-2$ at the expense of $N-3$
and so on. The last avoided crossing assumed to be passed diabatically
populates $n=1$ state and leaves empty $n=0$. 
It is clear that, provided all $N$ avoided crossings are passed
diabatically (all $N$ particles are successively excited),
 we realize the desired, described above final Slater
determinant involving the oscillator functions from
$n=1$ to $n=N$ as shown in Fig.~\ref{ideal}b.

We have shown \cite{kark01,bodzio01} that a similar laser sweeping scheme 
works very efficiently in the bosonic case. There, however, a single
avoided crossing (between the ground and the first excited state) had
to be passed diabatically by $N$ bosons.
In the present situation we have to cross diabatically
$N$ different avoided crossings.
 
Apart from the experimental approach which is obviously
beyond the scope of the present letter, the other way to test the
proposed scheme is the numerical integration of the time-dependent 
Schr\"odinger equation. 
Since we deal with $N$
noninteracting particles the problem of integration of $N$ dimensional
Schr\"odinger equation reduces to the problem of integration of
$N$ independent single particle 
equations for time evolution of the wave functions $\psi_i$, see
(\ref{slater}). 
Such an approach, with the
parameters assumed above [i.e. $\sigma$, $U_0(x_0)$ and $x_0(t)$] 
yields to transfer of particle from $n$  level 
to   $n+1$ level
with the probability $p$ higher than 99.95 \%, as we have 
checked for $n \leq 180$.  
This indicates that the proposed mechanism of excitation is extremely
efficient and to the end of this paper we will show analytical results 
assuming $p = 1$.

So far we have analyzed the system at the zero temperature limit. 
The lowest attainable temperature of a trapped Fermi 
gas in the recent experiments 
corresponded to $0.2T_F$ \cite{schreck01,truscott01}, 
where $T_F$ is the Fermi temperature.
To consider the finite temperature effect on the excitation process 
we switch to the 3D model of noninteracting fermions.
We assume a cigar-shaped harmonic trap with an additional local gaussian 
well created by the laser beam
\begin{eqnarray}
V(x,y,z) & = & \frac{x^2}{2} + \frac{\omega_\perp}{2 \omega} \ (y^2 + z^2)
\nonumber \\ 
&& +   U_0(x_0)
\exp\left(\frac{-(x-x_0)^2}{2\sigma^2}\right),
\label{potencja1}
\end{eqnarray}
where we have chosen $\omega_\perp/\omega = 10$ as an example.
The other parameters of the potential are exactly the same as 
in the 1D case. The laser beam is directed along 
$z$ axis. Its crosssection in the $xy$ plane 
is assumed to be tightly focused in the $x$ direction having at the same time
large waist along $y$. In effect
the gaussian character in the potential 
is realized in the $x$ direction only, while in the $y$ 
direction the laser intensity remains constant on the size of the Fermi cloud.
Changing the trapping potential by  sweeping of the laser beam
 allows for excitations 
in the $x$ degree of freedom of the system. The excitation in this degree 
is exactly the same as in the 1D case considered previously because
the 3D model of noninteracting particle is separable.

\begin{figure}
\centering
{\epsfig{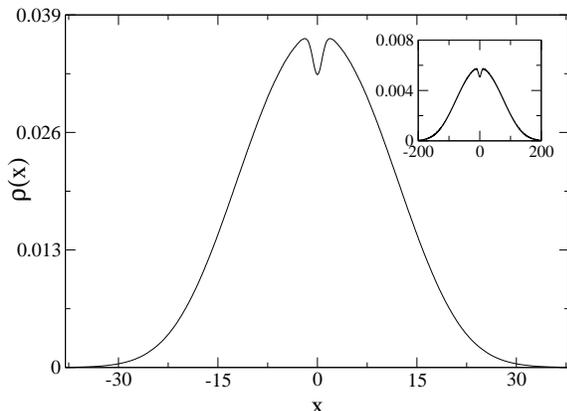}}
\caption{
Single particle reduced probability density after the excitation
process for $N = 40000$ fermions. 
The figure represents average over 100 samples 
corresponding to the Fermi-Dirac distribution with $T = 0.3 T_{F} $. 
The inset shows the same density after a free expansion 
lasting one period of the harmonic trap ($2\pi/\omega$). In the figure we use 
the harmonic oscillator unit of length, i.e. $\sqrt{\hbar/m\omega}$.
}
\label{temp}
\end{figure}

Assume we have $N$ fermions in the harmonic trap distributed among the 
 energy levels according to the Fermi-Dirac statistics with $T>0$. The
corresponding single particle probability density (integrated over $y$ and $z$ 
coordinates) reads
\begin{equation}
\varrho(x) = \frac{1}{N} \sum_{n_x} g_{n_x} |\psi_{n_x}(x)|^2,
\label{gestosc}
\end{equation}
where the sum runs over occupied $\psi_{n_x}$ states only. The degeneracy 
factor  $g_{n_x}$ takes into
account that several fermions may share the same $\psi_{n_x}$ state with
different other quantum numbers (basically $\sum_{n_x}g_{n_x}=N$).
Next we perform the potential sweeping that realizes excitation in the $x$  
degree of freedom, i.e. each occupied $\psi_{n_x}$ state goes to 
$\psi_{n_x+1}$ with a probability practically equal to unity.
In Fig.~\ref{temp} we show the results of the excitation process for 
$N=40 000$ fermions in the temperature $T=0.3T_F$. For example that corresponds to
$T=0.2\ \mu$K for $\omega = 2\pi\cdot \ 50$~Hz being within
 the range of current experiments 
\cite{demarco99,schreck01,truscott01}.
The results presented in Fig.~\ref{temp} were obtained by averaging over 
100 realizations of $N$ fermions in the temperature 
$T=0.3T_F$. A successful excitation results in a central dip
 clearly visible in the 
figure.
Note also that in the present case we can employ much more particles than 
in the 1D model and the dip in the probability density possesses still 
a considerable depth. That is due to the simple fact that
 before the potential sweeping
the $\psi_{n_x}$ state with $n_x=0$ contributes many ($g_0$) times to 
the density (\ref{gestosc}). Consequently, after the excitation, 
the lack of the $n_x=0$ state is much more significant than in the 
1D case where such a state could be occupied once only.

As a difficulty, however, one might realize that the
width of the dip is of the order of a few percent of the whole 
distribution of atoms in a harmonic trap.
This might preclude its direct observation. To
overcome this difficulty one may, for the detection purposes, supplement
the proposed scheme by the ``free fall expansion''. Assume that after
the sweep by the laser beam and the corresponding creation of the
collective excitation we turn off the trap completely. The
group of fermions will fall in the gravitational field, the cloud will
expand simultaneously. It is important to realize that the corresponding 
single particle density 
also expands roughly preserving its shape (and in particular
the central dip structure).  
The single particle density after the free expansion lasting 
$2\pi/\omega$ (i.e. 20~ms for $\omega = 2\pi\cdot \ 50$~Hz) is
shown in the inset of Fig.~\ref{temp}.
Observe that the width of the dip can quickly reach the size comparable to
the initial size of the whole cloud.

To summarize, we have presented a simple method that allows for creation of
collective excitation in a trapped Fermi gas in analogy to a soliton-like state
of a Bose-Einstein condensate. The results shown in this work are based on the
noninteracting particle model. We believe, that in analogy to the bosonic 
case, an interaction between fermions (much weaker than in the bosonic case) 
will not affect the efficiency of the proposed scheme.

The method can be also applied to create collective excitation 
in the so-called Tonks gases (see, e.g., \cite{tonks} and references therein). 
Indeed, impenetrable Bosons trapped in a quasi 1D potential can 
be modeled by a noninteracting Fermi gas. Hence, the excitation scheme
presented here becomes directly applicable to such a Bose system.

We are grateful to M.~Brewczyk, K.~Rz\c{a}\.zewski and L.~Santos for the
discussion. Support of KBN under project 5~P03B~088~21 is acknowledged.

\end{multicols}
\end{document}